**RESEARCH ARTICLE**

# A Benchmarking Dataset with 2440 Organic Molecules for Volume Distribution at Steady State

Wenwen Liu[1], Cheng Luo[2,3], , Hecheng Wang[1,*], Fanwang Meng[4,*]

[1]School of Life and Pharmaceutical Sciences, Dalian University of Technology, Panjin, 124221, China;

[2]University of Chinese Academy of Sciences, Beijing, 100049, China;

[3]Drug Discovery and Design Center, the Center for Chemical Biology, State Key Laboratory of Drug Research, Shanghai Institute of Materia Medica, Chinese Academy of Sciences, Shanghai, 201203 China;

[4]Department of Chemistry and Chemical Biology, McMaster University, Hamilton, Canada, L8S 4L8;

**Abstract: Background:** The volume of distribution at steady state ($VD_{ss}$) is a fundamental pharmacokinetics (PK) property of drugs, which measures how effectively a drug molecule is distributed throughout the body. Along with the clearance (CL), it determines the half-life and, therefore, the drug dosing interval. However, the molecular data size limits the generalizability of the reported machine learning models.

***Objective:*** This study aims to provide a clean and comprehensive dataset for human $VD_{ss}$ as the benchmarking data source, fostering and benefiting future predictive studies. Moreover, several predictive models were also built with machine learning regression algorithms.

**Methods:** The dataset was curated from 13 publicly accessible data sources and the DrugBank database entirely from intravenous drug administration and then underwent extensive data cleaning. The molecular descriptors were calculated with Mordred, and feature selection was conducted for constructing predictive models. Five machine learning methods were used to build regression models, grid search was used to optimize hyperparameters, and ten-fold cross-validation was used to evaluate the model.

**Results:** An enriched dataset of $VD_{ss}$ (https://github.com/da-wen-er/VDss) was constructed with 2440 molecules. Among the prediction models, the LightGBM model was the most stable and had the best internal prediction ability with $Q^2 = 0.837$, $R_{test}^2 = 0.814$, and for the other four models, $Q^2$ was higher than 0.79.

**Conclusions:** To the best of our knowledge, this is the largest dataset for $VD_{ss}$, which can be used as the benchmark for computational studies of $VD_{ss}$. Moreover, the regression models reported within this study can be of use for pharmacokinetic related studies.

## 1. INTRODUCTION

Drug research and development (R&D) is a complex and lengthy process, and according to statistics, about 90% of compounds entering clinical trials fail [1, 2]. Helen Dowden's analysis of the causes of clinical failures in 2016-2018 showed that 79% were safety or efficacy [3]. Highlighting the significance of pharmacokinetics (PK) in drug R&D [4, 5]. PK is a recognized fundamental property that affects the concentration of a drug at the target and ultimately determines the efficacy and safety [5, 6]. The volume of distribution at steady state ($VD_{ss}$) is one of the most crucial pharmacokinetic parameters because $VD_{ss}$ and clearance (CL) together determine the dosing frequency, and both help determine half-life ($t_{1/2}$) mean residence time (MRT) ($MRT = VD_{ss}/CL$) [7, 8]. It is defined as the ratio of its dose *in vivo* to its plasma concentration at equilibrium [9]:

$$VD_{ss} = \frac{A_{ss}}{C_{ss}} \qquad (1)$$

where $A_{ss}$ denotes the amount of drug in human body (blood phase) and $C_{ss}$ denotes plasma concentration at steady state. $VD_{ss}$ in humans can be obtained with many experimental methods [10, 11]. However, at present, the value of $VD_{ss}$ is still highly dependent on labor-intensive and costly *in vivo* and *in vitro* experiments, which places a heavy burden on the pharmaceutical industry [12]. Hence, researchers pay more and more attention to *in silico* quantitative structure-activity relationship (QSAR) methods to predict $VD_{ss}$ values as an alternative [12].

Applying QSAR models to predict human PK has been proved to be a useful strategy, potentially reducing R&D time, costs, and resources [13, 14] where the main idea of QSAR is to uncover the relationship between structural properties, e.g., the number of hydrogen bond donors (HBD) and acceptors (HBA), polar surface area, and $VD_{ss}$ with statistical models [15, 16]. Therefore, the results of QSAR studies can suggest which compounds should have a higher priority for further testing through more expensive but more accurate *in vivo* and *in vitro* experiments. Furthermore, the utterly computational method does not require any synthesis or animal experiments, only the computer calculation of descriptors and model building [17]. In addition to significantly reducing time and cost requirements, computational models have the advantage of being able to be generated directly from human data. They can even be used to evaluate the PK of compounds that have not yet been synthesized, which is impossible to do *in vivo* or *in vitro* experiments [18]. In short, these QSAR based models are less time-consuming and cost-effective than *in vivo* and *in vitro* methods.

Recently, machine learning (ML) algorithms have emerged as a powerful and flexible set of toolboxes for computational drug design [19-23], including $VD_{ss}$ [24-26]. A great benefit of ML algorithms is that they can uncover hidden patterns by extracting new knowledge from massive biomedical big data. Gleeson first used 199 compounds combined with three computational models of bayesian neural network (BNN), classification regression tree (CART), and partial least squares (PLS) to predict human $VD_{ss}$, and finally, combined the three methods to establish a CART-BNN-PLS model with $R_{test}^2 = 0.612$ [14].

In 2008, Obach published a database containing the main pharmacokinetic parameters of 670 drugs [27]. Giuliano *et al.* used this database in 2009 to use ML algorithms such as random forests (RF), PLS, multiple linear regression (MLR) combined with two descriptor calculation software MOE and VolSurf+, established MLR model with an accuracy rate squared cross validated correlation coefficient ($Q^2$) of 0.540 [28]. In 2012, Zhivkova *et al.* used the collected $VD_{ss}$ values of 132 acidic drugs to obtain an MLR model with $R_{test}^2$ of 0.687 [29]. Lombardo and Jing, in 2018, reported 1352 datasets covering detailed physicochemical, pharmacological, structural, and $VD_{ss}$ information of compounds [30]. In 2019, Wang used this data set to establish a prediction model using support vector machine (SVM), RF, gradient boosting machine (GBM), extreme gradient boosting XGBoost, and other methods where it was found that the prediction accuracy of SVM was the highest, $R_{test}^2$ of 0.870 and Q² of 0.77 [31].

Even many computational studies have been proposed, predictive models for $VD_{ss}$ have not been widely used in practice, which is limited by the poor generalizability of these ML models. The pre-trained ML models tend to give poor predictions when faced with unseen molecules, leading to poor generalizability. This is caused by the limited chemical diversity covered by the training molecular dataset [32, 33]. Therefore, a dataset with an enriched chemical diversity is in great need. Moreover, due to the lack of a benchmarking dataset for $VD_{ss}$ predictions, comparing the model performance between different computational models becomes a challenging task.

To migrate these problems, we aim to provide a large dataset to serve as the benchmark dataset for the incoming computational studies covering a broader chemical diversity. This is achieved by a well-design scheme of data curation of existing publicly available datasets of $VD_{ss}$. Secondly, attempts are also given to build predictive models for $VD_{ss}$ using ML algorithms. We also try to improve model interpretability to understand the relationships between chemical properties, as derived from structural attributes (i.e., computed descriptors), and human $VD_{ss}$.

To achieve the first objective, we collected the value of $VD_{ss}$ from a vast set of literature (13 peer-reviewed publications [7, 13, 14, 17, 27, 30, 31, 34-39] and DrugBank database (https://go.drugbank.com/) [40]). The final data was obtained after data collection, supplementation, deduplication, and outlier removal, the processing steps were freely available as source code in GitHub (https://github.com/da-wen-er/VDss), and all data were also freely available on GitHub. The physicochemical properties of the compounds in the analysis dataset were consistent with those of the published dataset, indicating that the dataset was effective and available. Furthermore, using this dataset for model building and optimization, the stability of the new model was higher than the published models.

## 2. MATERIALS AND METHODS

### 2.1. Dataset

#### 2.1.1. Data Collection and Selection

Data sources from 13 peer-reviewed literatures and DrugBank database were collected. The DrugBank database downloaded from the website (https://go.drugbank.com/) had much redundancy and could not be used directly, and the required information needed to be extracted first. Therefore, Kettle [41] filters were used for DrugBank data to extract information such as molecular name, the volume of distribution, and CAS number. Second, because the volume of distribution contained much irrelevant content, it must be manually selected. The following procedures were employed for data selection:

- Delete the data points whose volume of distribution was empty.

- All the units (L/kg, mL/kg, and L) were converted to L/kg by assuming the average human weight was 70 kg [30, 34, 38, 39].

- Only records with the volume of distribution were kept, and other records were deleted (e.g., oral distribution, apparent distribution).

- Healthy subjects and adults data were prioritized [7].

- When there were multiple data, used the range.

Finally, the literature data were combined with the remaining data from DrugBank into one initial data source as summarized in Table **1** (for all of the data, see the raw_data_all.xlsx in GitHub repository).

**Table 1.** Data source-related information.

| Year | Number of Instances | References |
|------|--------------------|------------|
| 2002 | 64 | 39 |
| 2004 | 121 | 38 |
| 2004 | 70 | 13 |
| 2006 | 207 | 14 |
| 2006 | 384 | 17 |
| 2008 | 670 | 27 |
| 2009 | 121 | 34 |
| 2013 | 569 | 37 |
| 2016 | 1130 | 7 |
| 2018 | 1352 | 30 |
| 2018 | 152 | 36 |
| 2019 | 1270 | 31 |
| 2019 | 1354 | 35 |
| 2020 | 778 | 40 |

### 2.1.2. Data Cleaning

The merged data had much redundant information and data cleaning was performed thereafter (Fig. (**1**)). Building a regression model requires accurate values, so first, delete the data with an empty $VD_{ss}$, and filter out the data with a range of $VD_{ss}$. Next, the duplicated items were removed by comparing the international chemical identifier (InChI) (an international chemical identification system) because the InChI strings were unique molecular identifiers [42]. In order to assign an InChI string for each molecule, SMILES representations were required. Therefore, CAS number and compound name were used to parse isomeric SMILES with PubChemPy (https://github.com/mcs07/PubChemPy), resulting in 7473 data samples. Moreover, all the salts were stripped out and molecules containing any metal atoms were dismissed. There were 2865 molecules remaining after removing the duplicated records by comparing the InChI strings. All the molecular operations were done with RDKit (http://www.rdkit.org/). Once the dataset was curated, data cleaning was conducted. Checked the information contained $logVD_{ss}$, molecular weight (MW), logP of the compounds. The discovery and processing of outliers were achieved using boxplots combined with scatterplots and cumulative distribution functions, c.f. Fig. (**2**). Interquartile range (IQR) was used and only data points that fall into the range between the first quartile and third quartile were kept. It was seen from the box plot that the raw data contained a lot of abnormal data. Finally, 2440 molecules with valid $VD_{ss}$ values were obtained.

## 2.2. Feature Engineering

### 2.2.1. Feature Generation

Many structural and physicochemical properties of compounds are related to biological activity. In addition, quantitative descriptions of the molecular physical, chemical, and topological properties are essential for reliable QSAR models. Existing proposals for $VD_{ss}$ prediction model development are mainly based on molecular descriptors as models' independent variables [43-45]. All the hydrogen atoms were added before generating the 3D coordinates. Then the merck molecular force field 94 static (MMFF94s) [46] was used for geometry minimization. Mordred [47] was used to compute the molecular descriptors, removing data that made errors during the calculation, resulting in 2420 drug compounds which include 1613 two-dimensional (2D) and 213 three-dimensional (3D) structural descriptors. That's saying, we have a feature matrix $X \in \mathbb{R}^{2420 \times 1826}$ .

### 2.2.2. Feature Selection

Feature processing, also known as feature engineering, was required to build high-precision models [48]. Therefore, feature engineering can be considered an essential step in ML. First, deleted erroneous descriptors that could not be used for computation. The feature types generated by Mordred include float, object, and bool. The computer could only calculate the features of the float type, so the bool was converted to float. Descriptors appeared strings because of the compounds themselves and Mordred, so the total number of the string in the descriptor of type "object" was counted, and a critical point was selected. The feature was deleted if the number of strings was more significant than or equal to the critical point. Else, the data corresponding to the string was deleted. If the critical point was too small, the remaining features would be significantly reduced, and if the critical point was too large, the remaining data would be too small. After calculation, it was found that with 30 as the critical point, the number of deleted features and data was not too much, which was the most suitable, and finally, 2370×1443 data and features were obtained. Besides, the number of features should be as small as possible to avoid overfitting and wasting computing resources. Therefore, feature selection is needed, one of the techniques used for dimensionality reduction, which selects features relevant to labels and discards irrelevant and redundant features [49, 50]. This study used a random data split method to select the training set and used the following four different methods for feature selection:

- Remove Low Variance Features. Eliminate features with a variance value of 0, removing features with the same value.

- Remove High Correlation Features. Feature correlation removal was performed using Pearson correlation coefficient, and the parameter threshold was set to 0.95. That was, only one of the two features with a correlation higher than 0.95 will be retained.

- Remove Features with Uneven Distributions in the Training and Test Sets. A kernel density estimation (KDE) plot is a method of visualizing the distribution of observations in a dataset, similar to a histogram, that represents the data using a continuous probability density curve in one or more dimensions and smoothes the observations using a Gaussian kernel, producing a constant density estimate [51, 52]. In this paper, the uneven features were found and removed using KDE.

- Wrapper. The wrapper method is a greedy algorithm. It uses the algorithm model to train and evaluate the feature subset and the target (label) set and measure the quality of the feature subset according to the training accuracy (accuracy rate) to select the best feature set [53, 54]. A recursive feature elimination (RFE) wrapper with the base model XGBoost was used for the final feature selection. RFE recursively selects features based on the score of each feature by comparing the current training model with the previous model [55]. XGBoost parameters of the base model were set as follows: random_state=42 to ensure the repeatability of the algorithm. Step=50 in RFE was used to specify the number of unimportant features removed in each iteration, and other parameters were default values.

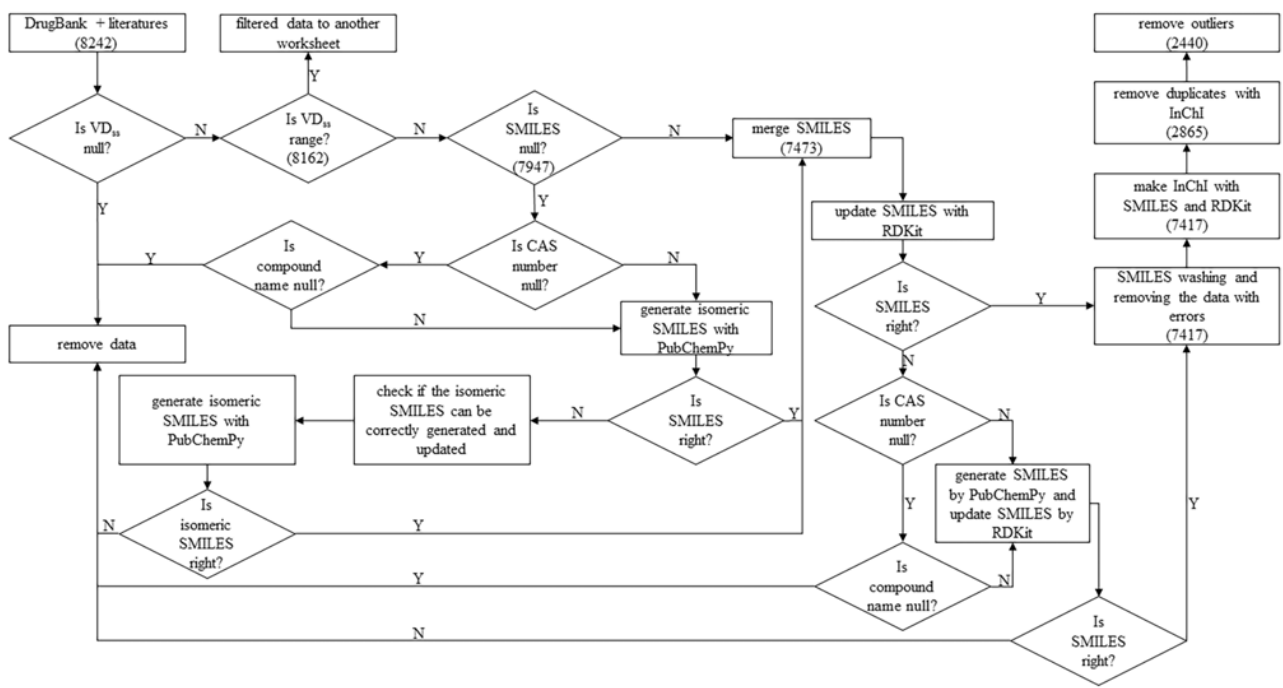

**Fig. (1).** Data cleaning process. First, the erroneous data from VD$_{ss}$ were deleted, followed by the supplementation and unification of the isomeric SMILES format, followed by the generation of InChI using isomeric SMILES, and finally by the deduplication of the data using InChI.

### 2.3. Machine Learning Model

#### 2.3.1. Model Construction

Five ML regression models were built, including RF [56], light gradient boosting machine (LightGBM) [57], support vector machine regressor (SVR, regression prediction model for SVM), [58], XGBoost [59], and gaussian process regression (GPR) [60], these techniques were robust, efficient, and have been widely used in QSAR modeling [7, 28, 37]. The XGBoost model was built with its own algorithm library, and the rest of the models were all built with the scikit-learn [61] package in the Python programming language. RF is an ensemble learning technique that improves CART using training data samples and random feature selection in tree induction [56]. LightGBM is a boosted ensemble model that transforms coupled weak learners into a latent model based on GBDT proposed by Microsoft in 2017 [57, 62]. LightGBM enhances the ability of gradient boosted decision tree (GBDT) models to speed up and reduce memory consumption at runtime while maintaining high accuracy. Due to a large amount of data, the accuracy of traditional GBDT-based prediction models decreases, and the prediction speed drops significantly. LightGBM model employs histogram-based algorithms to mitigate the effects of high-dimensional data, speed up computation time, and prevent overfitting prediction systems [63]. SVM regression is a machine learning method based on the structural risk minimization principle in statistical learning theory and is widely used in compound ADME properties and protein structure studies [64, 65]. XGBoost stands for "Extreme Gradient Boosting" and is an advanced implementation of the GBM algorithm [59]. XGBoost, also known as the "regularization boosting" technique, adds a regularization term to the cost function to control the complexity of the model. GPR is a stochastic method based on statistical learning and Bayesian theory that measures the similarity between points using a kernel function (KF) to predict values for unknown problems from training data [66].

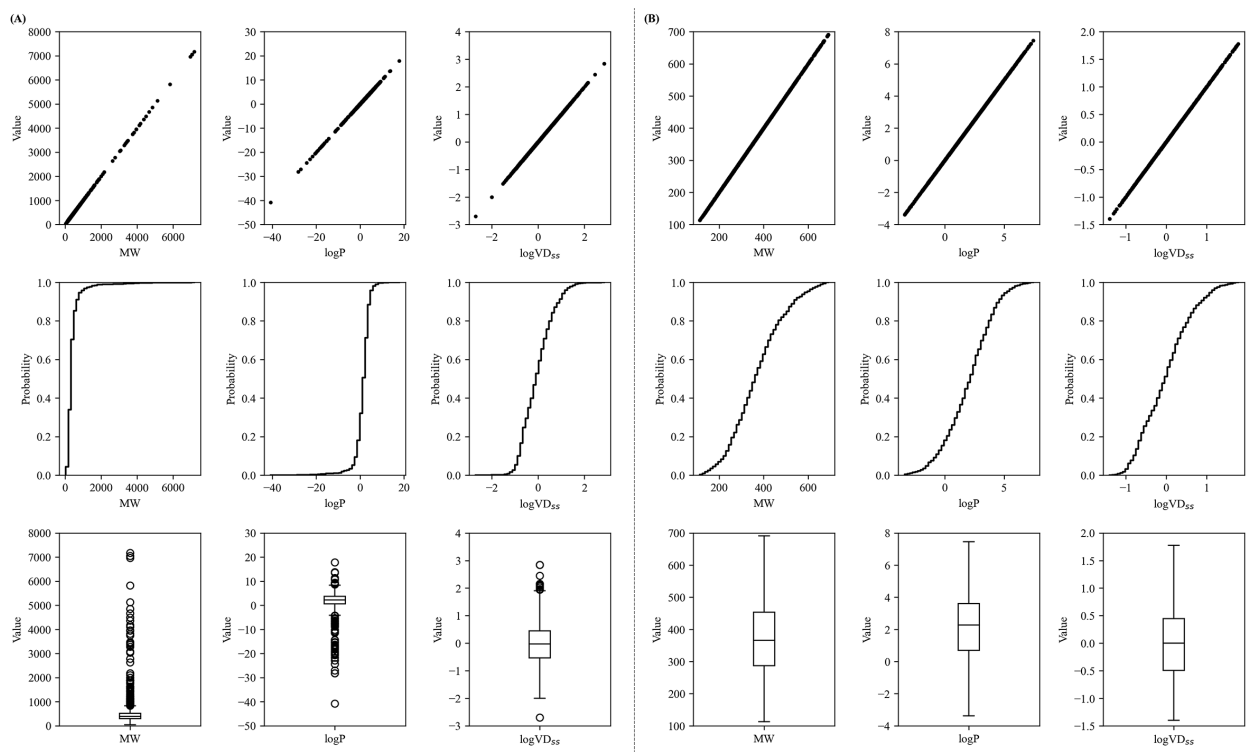

**Fig. (2).** Scatter plots, cumulative distribution function plots, and box plots before and after outlier processing. For each row in (A) and (B), scatter plots, cumulative distribution function plots, and box plots were listed for molecular weight, logP, and logVD$_{ss}$, respectively.

### 2.3.2. Model Optimization and Evaluation

The process of tuning parameters to bring the model to an ideal model is called hyperparameter optimization [67]. Optimization of hyperparameters is the key to building effective ML models, especially for tree-based ML models and deep neural networks with many hyperparameters [68]. In this study, the hyperparameters of the regression models were optimized with grid search. Grid search is an exhaustive search parameter adjustment method, which obtains the optimal parameter value by traversing parameters within a given range [69]. For RF, the range of n_estimators, max_depth, min_samples_leaf and main_samples_split were 270 to 310, 1 to 30, 1 to 11, and 2 to 22, respectively, step size was set to 1. For LightGBM, hyperparameters n_estimators and learning_rate were combined for optimization. The range of n_estimators was 280 to 305, the step size was set to 1, and the learning_rate was 10 values taken by numpy.linspace(0.01, 0.1, 10). The range of max_depth and num_leaves were set to 1 to 20, 2 to 50, respectively, the step size was set to 1. reg_alpha was 10 values taken by numpy.linspace(0.01, 0.1, 10). For SVR, the value of gamma and epsilon were 50 values and 10 taken by numpy.linspace(0.001, 0.05, 50) and numpy.linspace(0.001, 0.5, 10), respectively. The value range of C was 11 to 35, and the step size was set to 1. For XGBoost, hyperparameters n_estimators and learning_rate were combined for optimization. The range of n_estimators was 280 to 300, the step size was set to 1, and the learning_rate was 10 values taken by numpy.linspace(0.01, 0.1, 10). The range of max_depth and subsample all were 1 to 10, and the step size was set to 1. gamma was 20 values taken by numpy.linspace(0, 1, 20). For GPR, it was mainly the optimization of the kernel. In this article, the kernels were as follows:

a)   1.0 * RBF(length_scale=1) + WhiteKernel(noise_level=1)

b)   Matern(length_scale=0.484, nu=1.5) + WhiteKernel(noise_level=0.5)

c)   Matern(length_scale=0.484, nu=2.5) + WhiteKernel(noise_level=0.5)

d)   C(1.0, (1e-3, 1e3)) * RBF(10, (0.5, 2))

e)   DotProduct() + WhiteKernel(noise_level=0.5)

f)   Matern(length_scale=0.484, nu=1.5) + WhiteKernel(noise_level=1e-5)

g)   Matern(length_scale=0.484, nu=1.5) + WhiteKernel(noise_level=0.1)

h)   Matern(length_scale=0.484, nu=1.5) + WhiteKernel(noise_level=1)

i)   Matern(length_scale=0.01, nu=1.5) + WhiteKernel(noise_level=1)

j)   Matern(length_scale=0.1, nu=1.5) + WhiteKernel(noise_level=1)

k)   Matern(length_scale=0.5, nu=1.5) + WhiteKernel(noise_level=1)

l)   Matern(length_scale=1, nu=1.5) + WhiteKernel(noise_level=1)

Furthermore, alpha was 20 values taken by numpy.linspace(0, 1, 20).

The model performance was characterized with the coefficient of determination the square of the Pearson correlation coefficient ($R^2$) and the mean squared error (MSE), as defined:

$$R^2(y, \hat{y}) = 1 - \frac{\sum_{i=1}^{n}(y_i - \hat{y}_i)^2}{\sum_{i=1}^{n}(y_i - \bar{y})^2} \qquad (2)$$

$$MSE = \frac{1}{n}\sum_{i=1}^{n}(y_i - \hat{y}_i)^2 \qquad (3)$$

where n is the number of samples, $y_i$ are the experimental values, and $\hat{y}_i$ are the prediction values, and $\bar{y}$ are the mean values.

*2.3.3. Cross-Validation and Model Interpretation*

Ten-fold cross-validation was employed for model selection. Ten $R^2$ were obtained using equation (2), and the average value of $R^2$ was used to obtain $Q^2$, which was then used to evaluate the model. The larger the $Q^2$ value, the more stable the model and the better its internal prediction ability.

Analyzing the features used in the modeling process helps understand more about the model and the data. A single physicochemical descriptor alone cannot accurately predict $VD_{ss}$, but the effects of physicochemical properties on $VD_{ss}$ can help us understand the predictive models. In this study, feature importance was used to understand the relationship between model and features. Feature importance can be got in many ways. In this paper, the model XGBoost was used to evaluate the importance of 141 features. The "feature_importances" function was mainly used in scikit-learn to get the corresponding score. In addition, the correlation score was used to evaluate the degree of correlation between the feature and $VD_{ss}$, and the function "corr" is mainly used to get the corresponding value (method="spearman").

## 3. RESULTS AND DISCUSSION

### 3.1. A Benchmarking Dataset of $VD_{ss}$

By curating 14 public accessible datasets (including DrugBank), we curated a dataset of $VD_{ss}$ with 2440 molecules. This is the largest dataset for $VD_{ss}$, to the best of our knowledge. The $VD_{ss}$ values in the dataset ranged from 0.04 L/kg (Suprofen) to 60 L/kg (Amiodarone), and valid data points with MW values ranged from 110 to 700, logP values varied from -3.7 to 7.5, and logVD$_{ss}$ values varied from -1.5 to 1.7 were obtained.

The distributions of $VD_{ss}$ and logVD$_{ss}$ were shown in Fig. **3**(A) and (B). As shown in Fig. **3**(A), the distribution of $VD_{ss}$, 219 (90%) of the 2440 data had $VD_{ss}$ values between 0.1 and 10 L/kg; 998 data (41%) had a $VD_{ss}$ value less than or equal to 0.7 L /kg. The compounds of 8% had $VD_{ss}$ values greater than or equaled to 10 L/kg, and the larger the $VD_{ss}$ value, the more widely the drug was distributed in the human body. Fig. **3**(B) showed that the mean and median logVD$_{ss}$ were 0.03 and 0.0, respectively. The logVD$_{ss}$ values of 2199 (90%) compounds were all clustered between -1 and 1.

In addition, The collected 2440 molecules had extensive computational physicochemical properties. Fig. **3**(C) showed the distribution of MW, ranging from in which 2027 (83%) MW ranged from 110 to 500, which roughly met the MW limits set by the famous Lipinski rule of five, and the mean and median molecular weights were 376 and 366, respectively. logP expresses lipophilicity. Fig. **3**(D) showed the distribution of data logP, the mean and median were 2.1 and 2.3, respectively. The mean and median of HBD and HBA were 2.1 and 2.3, 5 and 5.5, respectively, histogram distribution as shown in Fig. **3**(E) and (F). The average values of logP, HBD, and HBA were all less than the limits set by the famous Lipinski rule of five (5, 5, and 10) [70]. The HBD and HBA values of 2199 (90%) compounds were less than 5. The median and mean rotatable bonds (RB) were 5 and 5.2, lower than the upper limit of 10 published by Veber *et al.* [71]. The median and mean topological polar surface area (TPSA) were 87 and 93.5, respectively. The histogram distribution of RB and TPSA was shown in Fig. **3**(G) and (H), respectively.

### 3.2. Feature Selection

By removing redundant features, regression error can be reduced, and computation speed improved [72]. The number of features deleted by different feature selection methods was shown in Table **2**. A total of 1826 features were generated using Mordred, of which 383 were deleted due to errors, and 133 were low variance features, which should also be deleted. Of the remaining features, 614 features were removed due to correlations higher than 0.95, which may indicate that these features have roughly the same predictive effect on logVD$_{ss}$. The features VR1_A, nB, NssssSi, n12Ring, n7aRing and n12FaRing, were found very unevenly distributed in the distribution of training and test set by KDE, should be removed. Finally, 549 features were removed using the wrapper method. After feature selection, 141 features remain for constructing the model, and all of them have a dimension of 2.

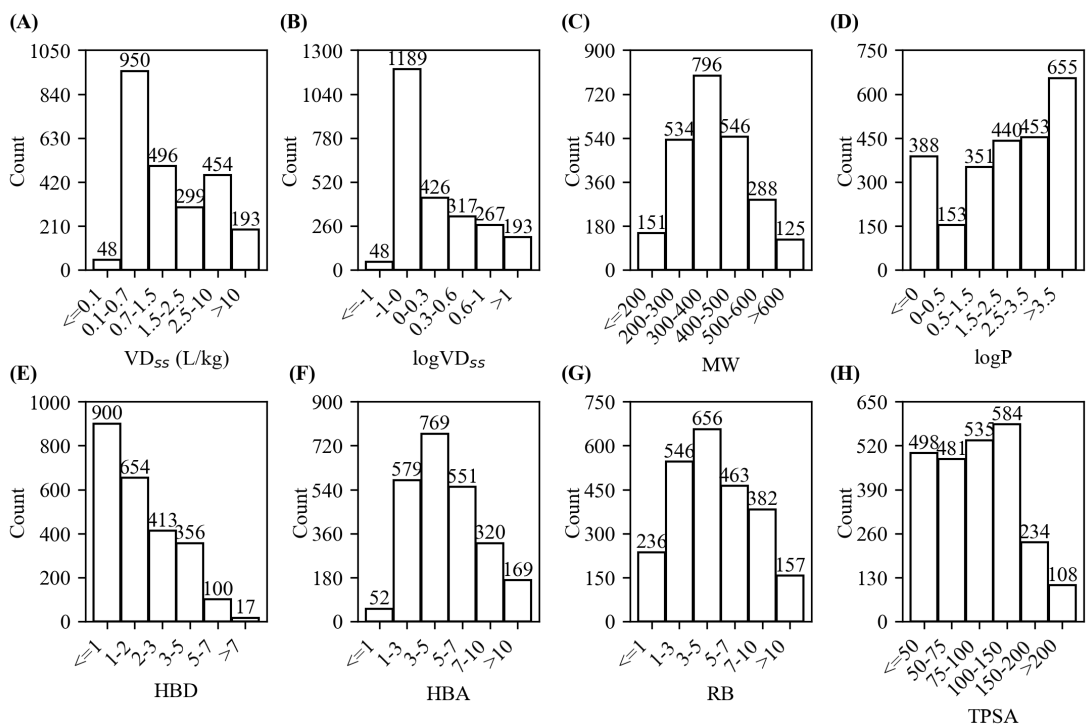

**Fig. (3).** Distribution of human pharmacokinetics values and computed physicochemical properties for the 2440 compounds. (A) VD$_{ss}$; (B) logVD$_{ss}$; (C) MW; (D) logP; (E) HBD; (F) HBA; (G) RB; (H) TPSA.

**Table 2.** The number of features to contain in the feature selection process.

|  | before Selection | Error Feature | Low Variance | High Correlation | Inconsistent Distribution | Wrapper | after Selection |
|---|---|---|---|---|---|---|---|
| Number of features | 1826 | 383 | 133 | 614 | 6 | 549 | 141 |

### 3.3. Model Result

#### 3.3.1. Model Construction and Optimization

Five machine learning models were built, namely RF, LightGBM, SVR, XGBoost, and GPR. Ten-fold cross validation was used for model selection purpose. Before hyperparameter optimization, the R$^2$ and MSE values of the five machine learning test sets were shown as test_original_R$^2$ and

test_original_MSE in Fig. (**4**), which suggested that XGBoost was the best performing model with $R^2_{test} = 0.809$ and $MSE_{test} = 0.075$ for the test set. The GPR model performs the worst with $R^2_{test}$ of 0.597 and $MSE_{test}$ of 0.159.

Grid search was used to optimize the model hyperparameters. In order to ensure the repeatability of the model, the parameter random_state was set to 42. The hyperparameters optimization results for the five models were: For RF model, the hyperparameter n_estimators=294. For LightGBM model, learning_rate=0.07, n_estimators=299. For SVR model, C=16, degree=1, epsilon=0.001, gamma=0.013. For the XGBoost model, the best results were obtained when n_estimators=295 and learning_rate=0.1. For GPR model, the best optimization result was obtained when kernel=Matern(length_scale=0.484, nu=1.5) + WhiteKernel(noise_level=1e-05). The hyperparameter optimization results not mentioned in the model were all default values. The optimization results of the five models were shown as test_optimization_R$^2$ and test_optimization_MSE in Fig. (**4**). Five models obtained large $R_{test}^2$ and low $MSE_{test}$ values, indicating that they effectively predicted human VD$_{ss}$. The LightGBM model had the best prediction effect, the $R_{test}^2$ was 0.814, and the $MSE_{test}$ was 0.073. XGBoost was the second most accurate predictor, with $R_{test}^2 = 0.812$ and $MSE_{test} = 0.074$. The SVR model had the lowest predictive power with test set scores of $R_{test}^2 = 0.768$ and $MSE_{test} = 0.091$. The RF model outperforms SVR slightly (RF: $R_{test}^2 = 0.782$ and $MSE_{test} = 0.086$). Hyperparameter optimization for GPR could improve model performance, with $R_{test}^2$ increased from 0.597 to 0.785 and $MSE_{test}$ decreased from 0.159 to 0.085.

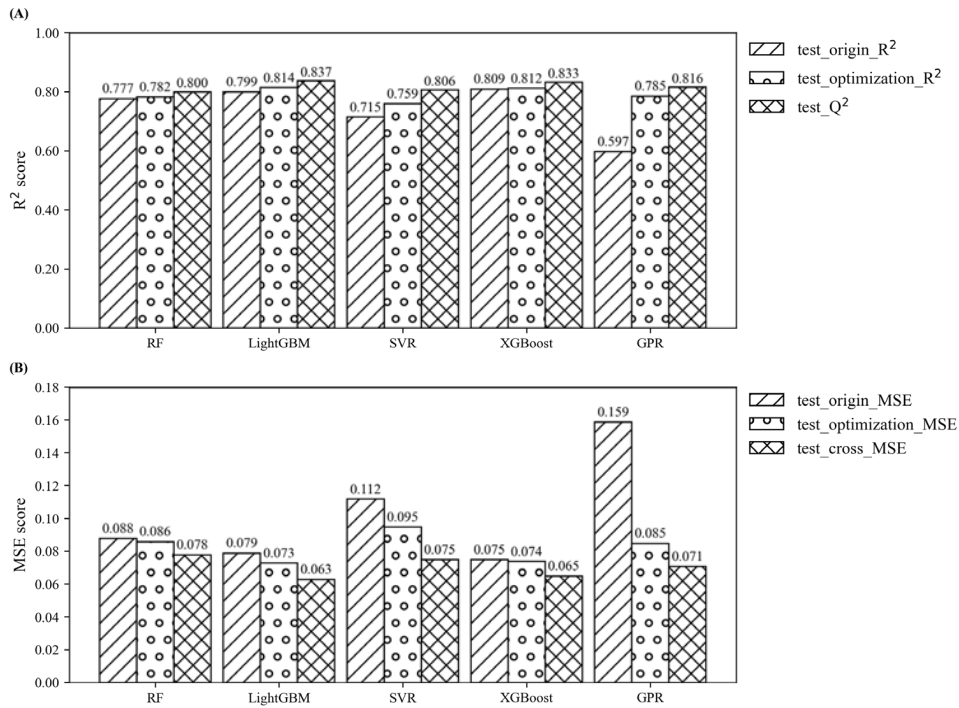

**Fig. (4).** The R$^2$ and MSE score of the models as built and after optimization and the model's Q$^2$ value. (A) R$^2$ score; (B) MSE score.

### 3.3.2. *Model Validation*

Ten-fold cross-validation was used to check the stability and predictive ability of the model. The Q$^2$ values of the five model test sets were shown as test_Q$^2$ and test_cross_MSE in Fig. (**4**), which were 0.800, 0.837, 0.806, 0.833, and 0.816, respectively, all greater than 0.5, indicating that these models had the specific predictive ability [26]. Moreover, LightGBM's Q$^2$ value was higher than the previous data published, and it was currently the model with the most stable prediction and the best internal prediction ability.

The relationship between the experimental logVD$_{ss}$ values and the predicted values for five different algorithms were shown in Fig. (**5**). It can be seen from the figure that the correlation coefficients were all greater than 0.85, of which

LightGBM was the largest with a correlation coefficient of 0.903, which indicates that the predicted value of logVD$_{ss}$ was relatively close to the actual value.

To further assess the predictability and stability of the developed model, the distributions of the cross-validation values of the five models concerning MSE, R$^2$, mean absolute error (MAE), and root mean square error (RMSE) were displayed in Fig. (**6**). It was illustrated in Fig. (**6**) that the LightGBM and GPR models were more robust than other models. In Fig. **6**(C), the MAE value for RF was the greatest, suggesting that the discrepancy between RF's anticipated and actual values was the greatest, and the model prediction's overall effect was the poorest.

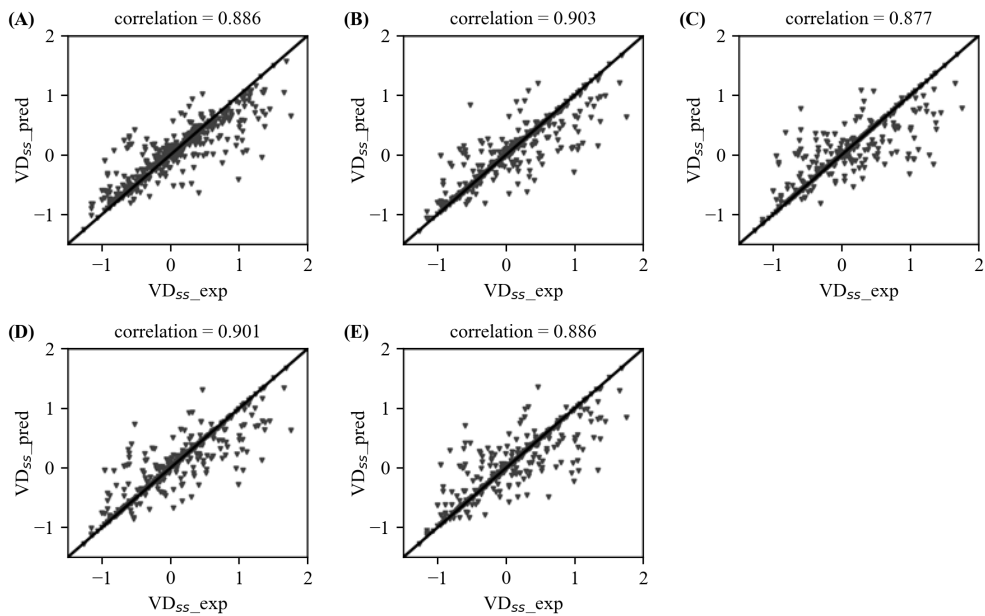

**Fig. (5).** Scatter distribution and correlation of predicted and experimental values for five models. (A) RF; (B) LightGBM; (C) SVR; (D) XGBoost; (E) GPR.

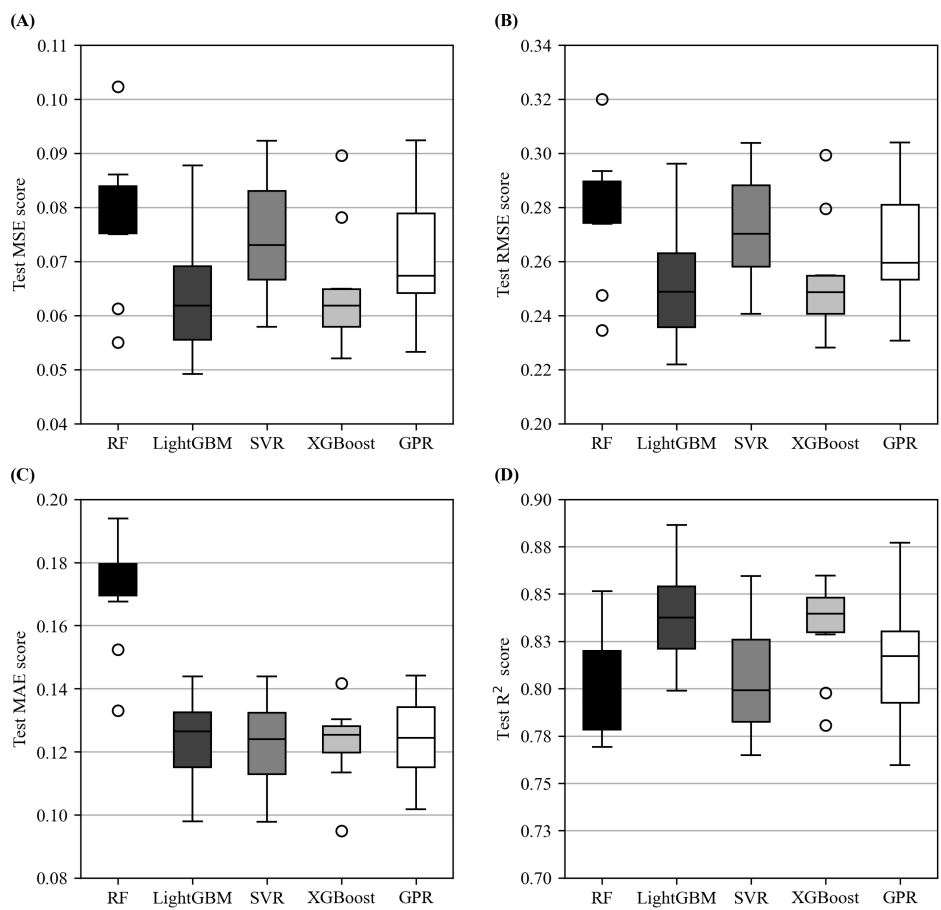

### 3.3.3. Model Interpretation

The feature information of the top ten feature importance scores among the 141 features were shown in Table **3**. These features were all 2D molecular descriptors. The correlation coefficient score between nAcid and logVD$_{ss}$ was -0.55, and the importance score was 0.088, which indicates that it had an enormous impact on logVD$_{ss}$ when modeling. Three features belonged to the autocorrelation module, namely GATS1p, AATS0P, AATS2p (their feature importance scores were 0.074, 0.024, 0.019, respectively), which may indicate that this module had a prominent contribution in predicting logVD$_{ss}$. In addition, nAcid and nBase belonged to the AcidBase model.

**Table 3.** Details of the ten features most relevant to VD$_{ss}$. (The basic descriptions of the features in the table are all from the website: https://mordred-descriptor.github.io/documentation/master/descriptors.html)

| Rank | Module | Name | Constructor | Dim | Description | Correlation Score | Importance Score |
|------|--------|------|-------------|-----|-------------|-------------------|------------------|
| 1. | AcidBase | nAcid | AcidicGroupCount () | 2D | acidic group count | -0.552092 | 0.087852 |
| 2. | Autocorrelation | GATS1p | GATS (1, 'p') | 2D | geary coefficient of lag 1 weighted by polarizability | 0.057177 | 0.073717 |
| 3. | TopoPSA | TopoPSA(NO) | TopoPSA (True) | 2D | topological polar surface area (use only nitrogen and oxygen) | -0.387521 | 0.025254 |
| 4. | Autocorrelation | AATSC0p | AATSC (0, 'p') | 2D | averaged and centered moreau-broto autocorrelation of lag 0 weighted by polarizability | -0.195637 | 0.024839 |
| 5. | SlogP | SlogP | SlogP () | 2D | Wildman-Crippen logP | 0.295047 | 0.024314 |
| 6. | MolecularId | AMID_O | 1.   MolecularId ('O', True, 1e-10) | 2D | averaged molecular ID on O atoms | -0.389201 | 0.023839 |
| 7. | AcidBase | nBase | BasicGroupCount () | 2D | basic group count | 0.325847 | 0.023785 |
| 8. | EState | NsssCH | AtomTypeEState ('count', 'sssCH') | 2D | number of sssCH | -0.079627 | 0.022317 |
| 9. | MoeType | SMR_VSA1 | SMR_VSA (1) | 2D | MOE MR VSA Descriptor 1 (-inf < x < 1.29) | -0.337933 | 0.022035 |
| 10. | Autocorrelation | AATS2p | AATS (2, 'p') | 2D | averaged moreau-broto autocorrelation of lag 2 weighted by polarizability | -0.154420 | 0.019331 |

## CONCLUSION

VD$_{ss}$ is an essential pharmacokinetic parameter, together with CL, determines half-life and thus dosing interval. Although VD$_{ss}$ can be obtained experimentally, it will consume many human resources, time, and money. These limitations further impede progress in drug discovery, and as a result, computational models to accelerate and reduce the cost of the drug R&D process are vigorously developed. However, current models' predictive ability and generalization ability remains to be further improved, which relying on extensive and accurate datasets with rich information. Therefore, a benchmarking dataset is in great need.

In this paper, we obtained a VD$_{ss}$ dataset containing 2440 human intravenous injections, which to our knowledge was currently the most extensive dataset on VD$_{ss}$, approximately twice the number of 1352 datasets published by Lombardo in 2018 [30], and the distribution of VD$_{ss}$ and related physical and chemical properties were unchanged. These data are valuable in studying the relationship between physical and chemical properties and VD$_{ss}$ and are available to researchers interested in the relationship between the human VD$_{ss}$ and structure. In addition, traditional experimental methods to obtain VD$_{ss}$ suffer from numerous limitations in time, cost, and resources. In this paper, using the collected 2370 compounds and 141 features for modeling and optimization, LightGBM stood out because of its best internal prediction capability with $R^2_{test}$ of 0.814, $MSE_{test}$ of 0.073, and Q² of 0.837. It can be concluded that our study is an essential application of ML models to human $VD_{ss}$ prediction and provides valuable guidance for early drug discovery. The dataset and related codes are free available at https://github.com/da-wen-er/VDss.

The accuracy of predictive models reported in this study can be furthered improved. Graph neural network (GNN) [73-75] based models have demonstrated their power in molecular property prediction with very satisfying performance and GNN will be used for $VD_{ss}$ predictions.

## LIST OF ABBREVIATIONS

| | | |
|---|---|---|
| BNN | = | Bayesian Neural Network |
| CART | = | Classification Regression Tree |
| GBDT | = | Gradient Boosted Decision Trees |
| GBM | = | Gradient Boosting Machine |
| CL | = | Clearance |
| GNN | = | Graph Neural Network |
| GPR | = | Gaussian Process Regressor |
| HBA | = | Hydrogen Bond Acceptor |
| HBD | = | Hydrogen Bond Donor |
| InChI | = | International Chemical Identifier |
| IQR | = | Interquartile Range |
| KDE | = | Kernel Density Estimate |
| KF | = | Kernel Function |
| LightGBM | = | Light Gradient Boosting Machine |
| MAE | = | Mean Absolute Error |
| ML | = | Machine Learning |
| MLR | = | Multiple Linear Regression |
| MMFF94s | = | Merck Molecular Force Field 94 Static |
| MRT | = | Mean Residence Time |
| MSE | = | Mean Square Error |
| MW | = | Molecular Weight |
| PLS | = | Partial Least Squares |
| PK | = | Pharmacokinetics |
| QSAR | = | Quantitative Structure Activity Relationship |
| Q² | = | Squared Cross Validated Correlation Coefficient |
| RB | = | Rotatable Bonds |
| RF | = | Random Forest |
| RFE | = | Recursive Feature Elimination |
| RMSE | = | Root Mean Square Error |
| R² | = | Squared Pearson's Correlation Coefficient |
| R&D | = | Research and Development |
| SVR | = | Support Vector Machine Regressor |
| SVM | = | Support Vector Machine |
| TPSA | = | Topological Polar Surface Area |
| $VD_{ss}$ | = | Volume of Distribution at Steady State |

## ETHICS APPROVAL AND CONSENT TO PARTICIPATE

Not applicable.

## HUMAN AND ANIMAL RIGHTS

No animals/humans were used for studies that are the basis of this research.

## CONSENT FOR PUBLICATION

Not applicable.

## AVAILABILITY OF DATA AND MATERIALS

All the codes and data for this study are free available at https://github.com/da-wen-er/VDss .

## FUNDING

None.

## CONFLICT OF INTEREST

The authors declare no conflict of interest, financial or otherwise.

## ACKNOWLEDGEMENTS

This work was financially supported by the Fundamental Research Funds for the Central Universities（DUT16RC(3)120）. Wenwen Liu, Cheng Luo, Hecheng Wang and Fanwang Meng initialized this project. Wenwen Liu performed data curation and data analysis. All the authors wrote the manuscript. We would like to acknowledge the scientific community for making the data sources publically available.